\documentclass[twocolumn,prl,aps,floats,showpacs]{revtex4}
\input epsf
\usepackage{amssymb}
\begin{document}
\title{Harmonic Measure and Winding of Conformally Invariant Curves}
\author{Bertrand Duplantier}
\affiliation{Service de Physique Th\'{e}orique de Saclay, F-91191
Gif-sur-Yvette Cedex, France}
\author{Ilia A. Binder}
\affiliation{Department of Mathematics, University of Illinois at Urbana-Champaign, IL 61801}

\date{August 2, 2002}

\begin{abstract}
The exact joint multifractal distribution  for the scaling
and winding of the electrostatic potential lines near any
conformally invariant scaling curve is derived in two dimensions.
Its spectrum $f(\alpha,\lambda)$ gives
 the Hausdorff dimension of the points where the potential scales with distance $r$ as $H \sim r^{\alpha}$
 while the curve logarithmically spirals with a rotation angle $\varphi=\lambda \ln r$. It obeys the scaling law
$f(\alpha,\lambda)=(1+\lambda^2)
f\left(\bar \alpha\right)-b\lambda^2$ with $\bar \alpha=\alpha/(1+\lambda^2)$ and
$b=(25-c)/{12}$, and where $f(\alpha)\equiv f(\alpha,0)$ is the pure
harmonic measure spectrum, and $c$ the conformal central
charge. The results apply to $O(N)$ and Potts models, as well as to
${\rm SLE}_{\kappa}$.
\end{abstract}
\maketitle
The geometric description of the random fractals
arising in Nature is a fascinating subject. Among these, the
study of the particular class of random clusters or fractal
curves arising in critical phenomena has led to fundamental
advances in mathematical physics. In {\it two dimensions} (2D), conformal field theory (CFT)
 has in particular demonstrated that statistical systems at their critical point
produce {\it   conformally invariant} (CI) fractal structures,
examples of which are the continuum scaling limits of random
walks (RW), i.e., Brownian motion, self-avoiding walks (SAW), and
critical Ising or Potts clusters. A wealth of exact methods has
been devised for their study: Coulomb gas, conformal invariance, and quantum
gravity methods \cite{nien,cardy,DK,duplantier4,BDprl5}. Recently, rigorous methods
have also been developed, with the introduction of the 
Stochastic L\"owner Evolution (${\rm  SLE}$) process, which mimics the
wandering of critical cluster boundaries, and gives a
probabilistic means of studying their random geometry \cite{schramm1}.

A refined way of accessing this random geometry is
provided by classical potential theory of 
electrostatic or diffusion field near such random fractal
boundaries, whose self-similarity is reflected in a {\it  
multifractal} (MF) spectrum describing the singularities of the
potential, also called the harmonic measure.
In 2D, the first exact examples appeared for the
universality class of random or self-avoiding walks, and
percolation clusters, which all possess
 the same harmonic MF spectrum \cite{BDprl23} (see also \cite{LW}), 
 in contradistinction to higher dimensions \cite{cates}. The general solution
 for the potential distribution near any CI fractal in 2D, obtained in \cite{BDprl5}, depends only
on the so-called {\it   central charge c}, the parameter labeling
the universality class of the underlying CFT (see also \cite{LSW,hast}). This solution can be
generalized to higher multifractal correlations, like the joint
distribution of potential on both sides of a
 simple scaling path \cite{BDjsp}. 

The important question remains of the {\it   geometry
 of the  equipotential lines} near a random (CI)
fractal curve. They are expected to wildly rotate, or wind, in a spiralling motion which closely follows
the boundary itself. The key geometrical object is the {\it  logarithmic spiral}, 
which is conformally invariant. The MF description should generalize to a
{\it   mixed} multifractal spectrum, accounting for {\it   both scaling
and winding} of the equipotentials \cite{binder}.

In this
Letter, we obtain the exact solution to this mixed MF spectrum
for any random CI curve. In particular, it is shown
to be related by a scaling law to the usual 
harmonic MF spectrum. We use conformal tools (fusing quantum
gravity and Coulomb gas methods), which allow the
description of Brownian paths interacting and winding with CI curves,
thereby providing a probabilistic description of the potential map.

{\it Harmonic Measure and Rotations}. Consider a single
(CI) critical random cluster, generically
called ${\cal C}$. Let $H\left( z\right) $ be the potential at the
exterior point $z \in {\rm  {\mathbb  C}}$, with Dirichlet boundary
conditions $H\left({w \in \partial \cal C}\right)=0$ on the outer
(simply connected) boundary $\partial \cal C$ of $\cal C$, and
$H(w)=1$ on a circle ``at $\infty$'', i.e., of a large radius
scaling like the average size $R$ of $ \cal C$. As is well-known,
$H\left( z\right)$ is identical to the probability that a Brownian path
starting at $z$ escapes to ``$\infty$'' without having hit
${\cal C}$. 

Let us now consider the {\it  degree with which the
curves wind in the complex plane about point} $w$ and call
$\varphi(z)={\rm  arg}\,(z-w)$. The multifractal formalism
\cite{bb,hent,frisch,halsey1}, here generalized to take into
account rotations \cite{binder}, characterizes subsets
${\partial\cal C}_{\alpha,\lambda}$ of boundary sites by a
H\"{o}lder exponent $\alpha$, and a rotation rate $\lambda$,
such that their potential lines respectively scale and {\it logarithmic spiral} as
\begin{eqnarray}
\nonumber
H\left( z \to w\in {\partial\cal C}_{\alpha,\lambda }\right) &\approx& r ^{\alpha },\\
\varphi\left( z \to w\in {\partial\cal C}_{\alpha,\lambda
}\right) &\approx& \lambda \ln\, r\ , \nonumber 
\end{eqnarray}
in the scaling limit $a_0 \ll r=|z-w| \ll R,$ where $a_0$ is the lattice mesh, if any. The Hausdorff dimension
${\rm  dim}\left({\partial\cal C}_{\alpha,\lambda }\right)=f\left(
\alpha, \lambda\right)$ defines
 the mixed MF spectrum, which is CI since {\it   under a conformal map
 both $\alpha$ and $\lambda$ are locally invariant}.

 Reversing the escaping Brownian path which represents the potential, one can also
consider the {\it harmonic measure} $H(w,r)$, which is the probability
that such a path starting at distance $R$ first hits the boundary in the
disk $B(w,r)$ of radius $r$ centered at $w \in \partial \cal C$,
and $\varphi (w,r)$ the associated winding angle of the path down
to distance $r$ from $w$. The {\it   mixed} moments of $H$ and
$e^{\varphi}$, averaged over all realizations of ${\cal C}$, are defined as
\begin{equation}
{\cal Z}_{n,p}=\left\langle \sum\limits_{w\in {\partial {\cal C}}_r} 
H^{n}\left(w,r\right) \exp\, (p\,\varphi (w,r)) \right\rangle
\approx \left( r/R\right) ^{\tau \left(
n,p\right) }, \label{Z}
\end{equation}
 where the sum runs over the centers of a covering of the boundary by disks of radius $r$, and where $n$ and $p$ are 
 real numbers. The scaling limit involves
multifractal scaling exponents $\tau \left( n,p\right)$
which vary in a non-linear way with $n$ and $p$ \cite{binder,bb,hent,frisch,halsey1}.
 They obey the symmetric double Legendre transform 
\begin{eqnarray}
\nonumber
\alpha &=&\frac{\partial\tau }{\partial n}\left( n,p\right) ,
\quad \lambda =\frac{\partial\tau }{\partial p}\left( n,p\right), \\ \nonumber
f\left( \alpha, \lambda \right)&=&\alpha n+\lambda p-\tau \left( n,p\right) ,\\
n&=&\frac{\partial f}{\partial \alpha}\left( \alpha, \lambda
\right) , \quad p=\frac{\partial f}{\partial \lambda }\left(
\alpha, \lambda \right).
\label{legendre}
\end{eqnarray}
Because of the ensemble average (\ref{Z}), values of $f\left(
\alpha,\lambda \right)$ can become negative for some domains of
$\alpha,\lambda$.

{\it   Exact Mixed Multifractal Spectra}. Each 2D conformally
invariant random statistical system
 can be labelled by its {\it   central charge} $c$, $c\leq 1$. Our main result is the following exact scaling law:
\begin{eqnarray}
\label{scalinglaw}
 f(\alpha,\lambda)&=&(1+\lambda^2) f\left(\frac{\alpha}{1+\lambda^2}\right)-b \lambda^2\ ,\\
\nonumber
 b&\equiv&\frac{25-c}{12}\geq 2\ ,
\end{eqnarray}
where $f\left({\alpha}\right)\equiv
f\left({\alpha},\lambda=0\right)$ is the usual harmonic
MF spectrum in the absence of prescribed winding, first
obtained in \cite{BDprl5}, which can be recast as:
\begin{eqnarray}
 f(\alpha)=\alpha+b-\frac{b\alpha^2}{2\alpha-1}\ .
\label{falpha}
\end{eqnarray}
We thus arrive at the very simple formula:
\begin{eqnarray}
 f(\alpha,\lambda)=\alpha+b-\frac{b\alpha^2}{2\alpha-1-\lambda^2}\ .
\label{falphalambda}
\end{eqnarray}
Notice that by conformal symmetry ${\sup}_{\lambda}f(\alpha,\lambda)=f(\alpha,\lambda=0)$, i.e., 
the most likely situation in the absence of prescribed rotation 
is the same as $\lambda=0$, i.e. {\it  winding-free}.
The domain of definition of the usual $f(\alpha)$ (\ref{falpha}) is $\alpha \geq
1/2$ \cite{BDprl5,Beur}, thus for $\lambda$-spiralling points Eq. (\ref{scalinglaw}) gives
\begin{eqnarray}
 {\alpha} \geq \frac{1}{2}({1+\lambda^2})\ ,
\label{alpha'}
\end{eqnarray}
in agreement with a theorem by Beurling \cite{Beur,binder}.

There is a geometrical meaning to the exponent $\alpha$. For an angle
 with opening $\theta$, $\alpha={\pi}/{\theta}$, thus the quantity ${\pi}/{\alpha}$ can be regarded as
 a local generalized angle with respect to the harmonic measure. The geometrical MF spectrum
of the boundary subset with such opening angle $\theta$ and spiralling rate
$\lambda$ reads from (\ref{falphalambda})
\begin{eqnarray}
\hat f(\theta,\lambda)\equiv f(\alpha=\frac{\pi}{\theta},\lambda)=\frac{\pi}{\theta}+b-b\frac{\pi}{2}
\left(\frac{1}{\theta}+
 \frac{1}{\frac{2\pi}{1+\lambda^2} -\theta}\right).
 \nonumber
\label{fthetalambda}
\end{eqnarray}

As in (\ref{alpha'}), the domain of definition in the
$\theta$ variable is $0 \le \theta \le \theta(\lambda)$, with $\theta(\lambda)={2\pi}/({1+\lambda^2})$. 
The maximum is reached when the two frontier strands about point $w$ locally collapse into a single 
$\lambda$-spiral, whose inner opening angle is $\theta(\lambda)$ \cite{Beur}.

In the absence of prescribed winding ($\lambda=0$), the maximum
$D_{\rm  EP}\equiv D_{\rm  EP}(0)={\sup}_{\alpha}f(\alpha,\lambda=0)$
gives the dimension of the {\it   external perimeter} of the fractal
cluster, which is a {\it simple} curve without double points, and may differ from the full hull \cite{BDprl5,DAA}.
Its dimension reads \cite{BDprl5}
$D_{\rm  EP}=\frac{1}{2}(1+b)-\frac{1}{2}\sqrt{b(b-2)}$.
This corresponds to typical values $\hat
\alpha=\alpha(n=0,p=0)$ and $\hat\theta={\pi}/{\hat \alpha}=\pi(3-2D_{\rm  EP}).$

For spirals, the maximum value
$D_{\rm  EP}(\lambda)={\sup}_{\alpha}f(\alpha,\lambda)$ still
corresponds  in the Legendre transform (\ref{legendre}) to $n=0$,
and gives the dimension of the {\it   subset of the  external
perimeter made of logarithmic spirals of type $\lambda$}. Owing to (\ref{scalinglaw})
we immediately get
\begin{eqnarray}
D_{\rm  EP}(\lambda)=(1+\lambda^2)D_{\rm  EP} -b \lambda^2 \ .
\label{supf}
\end{eqnarray}
This corresponds to scaled typical values $\hat
\alpha(\lambda)=(1+\lambda^2)\hat \alpha$, and $\hat
\theta(\lambda)=\hat \theta/(1+\lambda^2).$ Since $b \geq 2$ and
$D_{\rm  EP} \leq 3/2$, the EP dimension decreases
with spiralling rate, in a simple parabolic way.
\begin{figure}[t]
\epsfxsize=8.5truecm{\centerline{\epsfbox{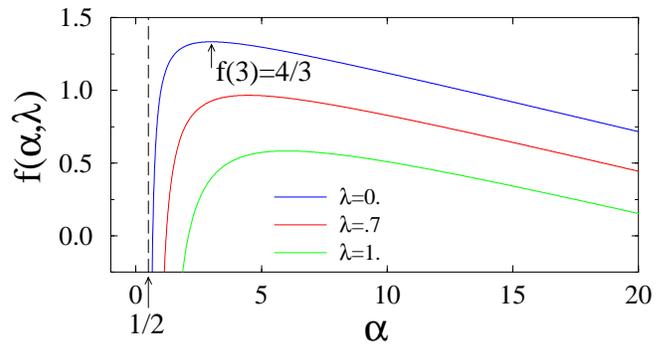}}}
\caption{Universal multifractal spectrum $f(\alpha,\lambda)$ for
$c=0$ (Brownian frontier, percolation EP and
SAW), and for three different values of the 
spiralling rate $\lambda$.} \label{Figure1}
\end{figure}
\begin{figure}[t]
\epsfxsize=8.5truecm{\centerline{\epsfbox{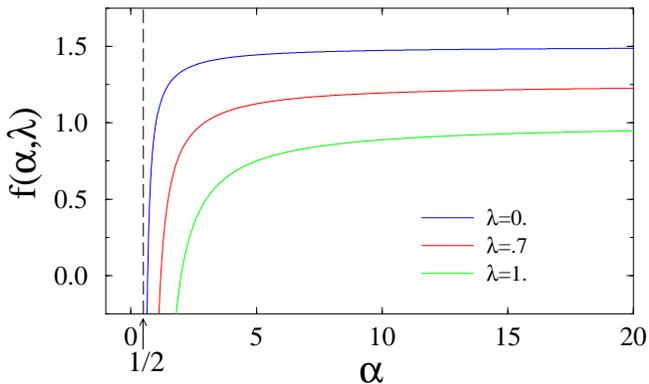}}}
\caption{Left-sided multifractal spectra $f(\alpha,\lambda)$ for
the limit case $c=1$ (frontier of a $Q=4$ Potts cluster or ${\rm  SLE}_{\kappa=4}$).}
\label{Figure2}
\end{figure}

Fig. 1  displays typical multifractal functions
$f(\alpha,\lambda\, ;c)$. The example choosen, $c=0$,  
corresponds to the cases of a SAW,
or of a percolation EP, the scaling limits of which both coincide
with the Brownian frontier \cite{BDprl23,LW}. The original
singularity at $\alpha=\frac{1}{2}$ in the
rotation free MF functions $f(\alpha,0)$, which describes boundary points with a needle local
geometry, is shifted for $\lambda \ne 0$ towards the minimal value (\ref{alpha'}). The right branch of
$f\left( \alpha,\lambda\right) $ has a linear asymptote
$\lim_{\alpha \rightarrow +\infty} f\left(\alpha,\lambda
\right)/{\alpha} =-(1-c)/24.$ Thus the $\lambda$-curves all become parallel for $\alpha \to
+\infty$, i.e., $\theta \to 0^{+}$, corresponding to deep fjords
where winding is easiest.

{\it   Limit} multifractal spectra are
obtained for $c=1$, which exhibit {\it   exact} examples of {\it  
left-sided} MF spectra, with a horizontal asymptote
$f\left(\alpha\to +\infty,\lambda\, ; c=1\right)=
\frac{3}{2}-\frac{1}{2}\lambda^2$ (Fig. 2). This corresponds to
the frontier of a $Q=4$ Potts cluster (i.e., the
${\rm  SLE}_{\kappa=4}$), a universal random scaling curve, with the
maximum value $D_{\rm  EP}=3/2$, and a vanishing typical opening
angle $\hat \theta=0$, i.e., the ``ultimate Norway'' where the EP
is dominated by ``{\it   fjords}'' everywhere
\cite{BDprl5,BDjsp}.

Fig. 3 displays the dimension $D_{\rm  
EP}(\lambda)$ as a function of the rotation
rate $\lambda$, for various values of $ c \leq 1$,
corresponding to different statistical systems. Again, the $c=1$
case shows the least decay with $\lambda$, as expected from the
predominence of fjords there.
\begin{figure}[t]
\epsfxsize=8.5truecm{\centerline{\epsfbox{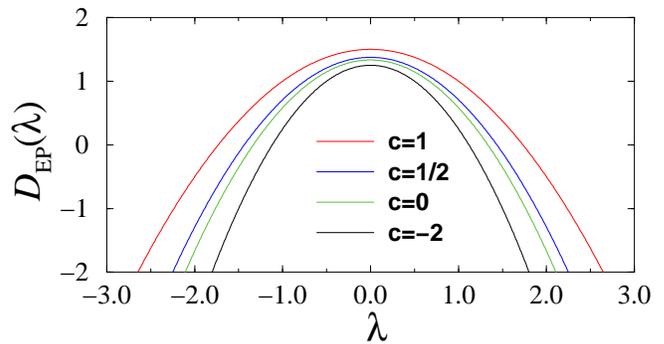}}}
\caption{Dimensions $D_{\rm  EP}(\lambda)$ of the external frontiers as a function of rotation rate.  The curves
are indexed by the central charge $c$, and correspond
respectively to: loop-erased RW ($c=-2\, ; {\rm SLE}_{2}$); Brownian or
percolation external frontiers, and self-avoiding walk ($c=0\, ; {\rm SLE}_{8/3}$);
Ising clusters ($c=\frac{1}{2}\, ; {\rm SLE}_{3}$); $Q=4$ Potts clusters ($c=1\, ; {\rm SLE}_{4}$).}
\label{Figure 3}
\end{figure}

{\it   Conformal Invariance and Quantum Gravity.} We
 now give the main lines of the derivation of exponents $\tau\left(
n,p\right)$, hence $f(\alpha,\lambda),$ by generalized {\it  
conformal invariance}. By definition of the $H$-measure, $n$ {\it  
independent} Brownian paths ${\cal B}$, starting a small distance $r$ away from
a point $w$ of the frontier $\partial \cal C$, and
diffusing without hitting $\partial \cal C$, give a geometric
representation of the $n^{th}$ moment$, H^{n},$ in Eq.(\ref{Z})
for $n$ {\it   integer}. Convexity yields analytic continuation for
arbitrary $n$'s. Let us introduce an abstract (conformal) field
operator $\Phi_{{\partial\cal C}\wedge {n}}$ characterizing the
presence of a vertex where $n$ such Brownian paths and the cluster's
frontier diffuse away from each other in a {\it   mutually-avoiding} configuration noted
${\partial\cal
C}\wedge {n}$ \cite{BDprl23}; to this operator is associated a scaling
dimension $x(n)$. To measure rotations as in moments (\ref{Z}) we have to consider
expectation values with insertion of the mixed operator
\begin{equation}
\Phi_{{\partial\cal C}\wedge n} e^{p\,{\rm  arg}({\partial\cal
C}\wedge n)} \longrightarrow x\left( n,p\right)=\tau \left( n,p\right)+2 , \label{xnp}
\end{equation}
where ${\rm  arg}({\partial\cal C}\wedge n)$ is the winding angle
common to the frontier and to the Brownian paths, and where  $x(n,p)$ is the
{\it scaling dimension}. One has $x(n,p=0)=x(n)$, and
$\tau \left( n,p=0\right)\equiv\tau \left( n\right)=x\left(
n\right) -2$.

Let us now use a fundamental mapping of the
CFT in the {\it   plane} ${\mathbb R}^{2}$  to the CFT on a
fluctuating abstract random Riemann surface, i.e., in presence of
{\it   2D quantum gravity} (QG) \cite{KPZ}. Two universal functions $U$
and $V$, acting on scaling dimensions, describe this map:
\begin{eqnarray}
U\left( x\right) &=&x\frac{x-\gamma}{1-\gamma} , \hskip2mm
V\left( x\right) =\frac{1}{4}\frac{x^{2}-\gamma^2}{1-\gamma}.
\label{UV}
\end{eqnarray}
with $V\left( x\right) \equiv U\left(
{\textstyle{1 \over 2}}%
\left( x+\gamma \right) \right)$\cite{BDprl23,BDprl5}. The parameter $\gamma$
 is the solution of
$c=1-6{\gamma}^2(1-\gamma)^{-1}, \gamma \leq 0.$

For the purely harmonic exponents $x(n)$,
describing the mutually-avoiding set ${\partial\cal C}\wedge n$, we have \cite{BDprl23,BDprl5}
\begin{eqnarray}
x(n)&=&2V\left[2 U^{-1}\left(
\tilde{x_1}\right)
 +U^{-1}\left( n \right) \right],  
\label{xcn}
\end{eqnarray}
where $U^{-1}\left( x\right) $ is the positive inverse of $U$
\begin{eqnarray}
\nonumber
2 U^{-1}\left( x\right)
=\sqrt{4(1-\gamma)x+\gamma^2}+\gamma\, .
\label{u1}
\end{eqnarray}
In (\ref{xcn}), the arguments $\tilde{x_1}$ and $n$ are respectively the {\it   boundary} scaling dimensions
(b.s.d.) of the simple path ${\cal S}_1$ representing a
semi-infinite random frontier (such that ${\partial\cal C}\equiv {\cal
S}_1\wedge{\cal S}_1$),
 and of the packet of $n$ Brownian paths, both diffusing into the upper {\it   half-plane} $\mathbb H$.
The function $U^{-1}$ maps these half-plane b.s.d's to the corresponding b.s.d.'s in quantum
gravity, the {\it linear combination} of which gives, still in QG, the
b.s.d. of the {\it   mutually-avoiding set} ${\partial\cal
C}\wedge n=(\wedge{\cal S}_1)^2\wedge n$. The function $V$ finally
maps the latter b.s.d. into the scaling dimension in $\mathbb R^2$.
The path b.s.d. $\tilde{x_1}$ obeys $U^{-1}\left(
\tilde{x_1}\right) =(1-\gamma)/2$ \cite{BDprl5}.

It is now useful to consider $k$ semi-infinite
random paths ${\cal S}_1$, joined at a single vertex in a
{\it   mutually-avoiding star} configuration ${\cal
S}_k=\stackrel{k}{\overbrace{{\cal S}_1\wedge{\cal
S}_1\wedge\cdots{\cal S}_1}}=(\wedge{\cal S}_1)^k$.
Its scaling dimension can be obtained from the same b.s.d. additivity rule in quantum gravity, as in
(\ref{xcn}) \cite{BDprl23,BDprl5}
\begin{eqnarray}
x({\cal S}_k)&=&2V\left[ k\, U^{-1}\left( \tilde{x_1}\right) \right]\ .  
\label{xk}
\end{eqnarray}
The scaling dimensions (\ref{xcn}) and (\ref{xk}) coincide when
\begin{equation}
x(n)=x({\cal S}_{k(n)}),\,\,\, k(n)=2+\frac{U^{-1}\left( n
\right)}{U^{-1}\left( \tilde{x_1}\right)}.
\label{kn}
\end{equation}
Thus we state the {\it   scaling star-equivalence}
${\partial\cal C}\wedge n \Longleftrightarrow {\cal S}_{k(n)}$,
{\it   of two simple paths ${\cal S}_1$ avoiding $n$ Brownian motions to $k(n)$
simple paths in a mutually-avoiding star configuration}, an
equivalence which will also play a essential role in the complete rotation
spectrum (\ref{xnp}).

{\it   Rotation scaling exponents}. The Gaussian distribution of
the winding angle about the {\it   extremity} of a scaling path,
like ${\cal S}_1$, was derived in \cite{DS}, using exact Coulomb
gas methods. The argument can be generalized to the winding angle
of a star ${\cal S}_k$ about its center \cite{BDtopub}, where
one finds that the angular variance is reduced by a factor
$1/k^2$ (see also \cite{wilson}). The
scaling dimension associated with the rotation scaling operator
$\Phi_{{\cal S}_k}e^{p\, {\rm  arg }\left({\cal S}_k\right)}$ is found
by analytic continuation of the Fourier transforms evaluated there \cite{BDtopub}:
\begin{eqnarray}
x({\cal S}_k;p)=x({\cal
S}_k)-\frac{2}{1-\gamma}\frac{p^2}{k^2}  \ , \nonumber 
\end{eqnarray}
i.e., is given by a quadratic shift in the star scaling exponent. To calculate
the scaling dimension (\ref{xnp}), it is 
sufficient to use the star-equivalence (\ref{kn}) above to
conclude that
\begin{eqnarray}
x(n,p)=x({\cal
S}_{k(n)};p)=x(n)-\frac{2}{1-\gamma}\frac{p^2}{k^2(n)}  \ ,
\nonumber
\end{eqnarray}
which is the key to our problem. Using Eqs (\ref{kn}),
(\ref{xcn}), and (\ref{UV}) gives the
useful identity:
\begin{eqnarray}
\frac{1}{8}({1-\gamma})k^2(n)=x(n)-2+b \ ,
\nonumber
\end{eqnarray}
with $b=\frac{1}{2}\frac{(2-\gamma)^2}{1-\gamma}=\frac{25-c}{12}$.
Recalling (\ref{xnp}), we arrive at the multifractal result:
\begin{eqnarray}
\tau(n,p)=\tau(n)-\frac{1}{4}\frac{p^2}{\tau(n)+b}  \ ,
\label{taunp}
\end{eqnarray}
where $\tau(n)=x(n)-2$ corresponds to the purely harmonic
spectrum with no prescribed rotation.

{\it   Legendre transform}. The structure of the full
$\tau$-function (\ref{taunp}) leads by a formal Legendre transform
(\ref{legendre}) directly to the identity
\begin{eqnarray}
 f(\alpha,\lambda)&=&(1+\lambda^2)f(\bar{\alpha})-b\lambda^2 \ ,
 \nonumber
\end{eqnarray}
where $f(\bar{\alpha})\equiv\bar{\alpha}n-\tau(n)$, with $\bar \alpha={d{\tau}(n)
}/{dn}$, is the purely
harmonic MF function. It depends on the natural reduced variable $\bar{\alpha}$
{\it   \`a la} Beurling ($\bar{\alpha} \in \left[\frac{1}{2},+\infty \right)$)
\begin{eqnarray}
\nonumber
\bar{\alpha}&\equiv&\frac{\alpha}{1+\lambda^2}=\frac{d{x}
}{dn}\left( n\right)=\frac{1}{2} +\frac{1}{2}
\sqrt{\frac{b}{2n+b-2}}\, ,
\end{eqnarray}
whose expression is found explicitly from (\ref{xcn}). Whence Eq.(\ref{scalinglaw}), {\bf QED}.

{\it   $O(N)$ and Potts models, ${\rm  SLE}_{\kappa}$}. Our results apply
to the critical $O(N)$ loop model, or to the EP's of critical Fortuin-Kasteleyn
(FK) clusters in the $Q$-Potts model, all described in terms
of Coulomb gas with some coupling constant $g$ \cite{nien}. ${\rm  SLE}_{\kappa}$ paths  also
describe cluster frontiers or hulls. One has the correspondence $\kappa={4}/{g}$, with a central charge
$c=(3-2g)(3-2g')=\frac{1}{4}({6-\kappa})(6-{\kappa'})$,
symmetric under the {\it duality} $gg'=1$ or $\kappa \kappa'=16$. This duality gives FK-EP's as 
some simple random $O(N)$ loops, or, equivalently, the ${\rm  SLE}_{\kappa'\leq 4}$ as the simple frontier of the
${\rm  SLE}_{\kappa \geq 4}$ \cite{BDprl5,BDjsp}.

\begin{acknowledgments}
The gracious hospitality of the Mittag-Leffler Institute, where
this work was initiated, is gratefully acknowledged, as is a
careful reading of the manuscript by T. C. Halsey.
\end{acknowledgments}

\end{document}